# Various Secure Routing Schemes for MANETs: A Survey

Priya R. Soni, Charmi A. Joshi, Dhwani R. Bhadra, Nikita P. Vyas, Rutvij H. Jhaveri

*Abstract*— MANET is an infrastructure less as well as self configuring network consisting of mobile nodes communicating with each other using radio medium. Its exclusive properties such as dynamic topology, decentralization, and wireless medium make MANET to become very unique network amongst other traditional networks, thereby determining security to be a major challenge. In this paper, we have carried out the survey of various security approaches of Mobile Adhoc Networks and provide a comprehensive study regarding it. We have focused our work on three approaches such as Bayesian watch dog, Trust based systems, and Ant colony optimization. In wireless perspective, security is a crucial term to handle. Therefore it becomes necessary when we are concerning our work with Mobile Adhoc Network.

## I. INTRODUCTION

Since the concept of computers and mobile technology is emerging, the need of wireless technology, a secure connection and a good communication range is been increased. Mobile Adhoc networks (MANET) are a kind of wireless networks that can usually change its location and is self configuring as well as self healing as they are mobile. The growth of laptop and 802.11/Wi-Fi wireless networking has made MANETs a popular research topic since the mid-1990s. There are various mobile devices in this network connected to each other via wireless medium. These devices are free to move anywhere in the fly and can communicate with each other stochastically at any place any time [1]. In this paper we have made a critical study on three main approaches to security which are trust based systems, Bayesian watchdog and ant colony optimization.

This characteristic makes MANET more vulnerable to be exploited by an attacker from inside the network. Wireless links also makes the MANET more susceptible to attacks which make it easier for the attacker to go inside the network and get access to the ongoing communication [3, 4]. Mobile nodes present within the range of wireless link can overhear and even participate in the network.

The trust based system firstly calculates the trust value using the information such as number of packets forwarded and number of packets to forward. Thus trust value is the ratio of number of packets receives to the number of packet sent and is dependent on the neighbors. This trust value is then compared with a threshold value and if it is found to be less, then the neighbor nodes corresponding to it are considered as malicious and thus the neighbor table is updated. Collaborative Bayesian technique proves to be more efficient and accurate than standard watchdog technique [6]. ACO algorithms are inspired by a foraging behavior of group of ants which are able to find optimal path based upon some defined metric which is evaluated during the motion of ants. ACO routing algorithms use simple agents called artificial ants which establish optimum paths between source and destination that communicate indirectly with each other by means of stigmergy.

Rest of the paper is organized as follows: In section 2 theoretical background regarding security issues in Manet and basic information about various security approaches are described. In section 3 related works on various security approaches is mentioned and in final section paper is concluded.

## II. THEORETICAL BACKGROUND

In the last few years, security of computer networks has been one of the serious issues which have widely been discussed. With the emergence of ongoing and new approaches for networking, new problems and issues arises for the basics of routing. Due to various factors like lack of infrastructure, absence of already established trust relationship in between the different nodes and dynamic topology, the routing protocols are vulnerable to various attacks [2]. Major vulnerabilities which have been researched till date are mostly these types which include selfishness, dynamic nature, and severe resource restriction and also open network medium. Despite of the above mentioned protocols in MANET, there are attacks which can be categorized in Passive, Active, Internal, External and network-layer attacks, Routing attacks and Packet forwarding attacks.

*A. Some Popular Approaches*

Bayesian Watchdog acts as intrusion detection system which analysis and overhears network traffic to detect misbehavior. Thereafter it computes trust value for each neighbor. If node's trust level is beyond certain threshold than node is considered as malicious. By analyzing second hand information using collaborative baysian watchdog, performance of system increases resulting in reducing amount of false negatives and increasing detection speed..[6]

While the routing in MANET is carried out using trust value hence forth trust based system is used. If the trust value is below the trust threshold the corresponding intermediate node is marked as malicious and isolated from the network, thus by increasing the performance of the secure network routing.

### III. RELATED WORK

We have carried out literature survey of the three approaches which are described as follows:

*1) Bayesian Watchdog:*

Manuel D. *et al.* [6] proposed that by analyzing second hand information using collaborative Bayesian watchdog, performance of system increases resulting in reducing amount of false negatives and increasing detection speed. In Bayesian watchdog techniques, Bayesian filter probabilistically determines state of system which is further pass over to beta function for estimating malicious nodes

Serrat-Olmos *et al.* [22] proposed Integration of various watchdogs that is collaborative watchdog. By using Collaborative watchdog approach, the detection time of misbehaved nodes is reduced and the overall accuracy is increased. It is concluded that a Bayesian watchdog performs better than a standard watchdog, by reducing the amount of false positives.

Sergio Marti *et al.* [23] proposed Classification of nodes based upon their behavior measured dynamically .Two techniques were used that is watchdog and path rater. Watchdog identifies malicious nodes while path rater helps routing protocols to avoid these misbehaving nodes to disturb network. It resulted in increasing throughput in network by keeping mobility moderate.

Jorge Hortelano *et al.* [24] proposed an approach for detecting black-hole attacks and selfish nodes in mobile P2P networks by using a watchdog sensor and a Bayesian filtering .In this Bayesian filters determines state of system probabilistically. Comparison of Bayesian and standard watchdog is made to test whether Bayesian filters works to support in accuracy to detect malicious nodes or not.

Varsha Himthani *et al.* [25] proposed a technique to integrate selfishness information with Bayesian filtering. It resulted in decrease in number of false positives. It aimed at increasing accuracy of detection of selfish and malicious nodes in MANET. By adding selfishness information with Bayesian filters leads in optimizing efficiency of watchdog intrusion detection system in peer to peer networks as MANET.

Nidal Nasser *et al.* [26] proposed an intrusion detection system which is extension of watchdog that is Ex-watchdog to remove weakness of Watchdog. It solved the problem of detecting a malicious node which can partition the network by falsely reporting other nodes as misbehaving inspite of misbehaving itself. It functions by detecting intrusion from malicious nodes and reports this information to the response system which is Path rater or route guard.

*2) Trust Based systems:*

Jing-Wei Huang *et al.* [15] proposed a security message scheme for MANETs which is based on trust based AOMDV. The process is carried out in four stages: Deciding the message and path degree of secrecy, encryption of message, routing and decryption. Hence by ns2 simulations it is observed to be more secured than T-DSR.

X. Li Z. *et al.* [16] proposed use of AOTDV protocol where a source can establish multiple-loop free paths to destination in one route discovery choosing the shortest path from it. After that the node trust values and number of hops are obtained, also the packet forwarding ratio is calculated. Packet forwarding ratio is used to analyze quality of forwarding. Here number of hops as well as the trust values to the path destination is evaluated

Partha Sarathi. *et al.* [17] proposed the use of AODV protocol with the use of fuzzy logic. Various parameters such as Reliability, Residual Energy, Buffer Occupancy, Packet Generation rate and Speed of the node are obtained for trust nodes, later these values are sent to fuzzy rule based trust calculation module which gives corresponding trust values.

N. Marchang *et al.* [18] proposed AODV protocol in which it selects most trusted path rather than the shortest path. Evaluations of trust values are carried out wherein is categorized further in self trust and neighbor trust. The main aim here is to make the protocol light weight.

Sherin Zafar *et al.* [19] proposed the use of trust based Qos i.e. TBQP protocol is done which also includes genetic algorithm as well. Firstly the fittest shortest path is obtained by using GA algorithm.

Na Li *et al.* [20] proposed opportunistic networks where PFM (Positive forwarding message) is said to be the evidence of the forwarding behavior of the node is considered. Besides it is consists of trust based framework having three modules-reputation modules, trust evaluation module and forwarding decision module.

*3) Bio-Inspired Approaches:*

There are several bio-inspired approaches; here we have studied the Ant Colony Optimization.

Jianping Wang *et al.* [7] proposed HOPNET which chooses the best shortest path from source to destination for this, the ant first select the node which is not visited yet by other ants after that it find the destination node in the routing table by exploring the adjacent link node. If the ant finds any unvisited node, then first it chooses that otherwise it chooses the node by pheromone concentration.

Mitthias Strobbe *et al.* [8] proposed in the ANTNET algorithm, there are two types of ants consists forward ant and backward ant. Forward ant firstly finds the shortest path from source to destination. If the ants successfully reach at the destination then remaining ant will follow the same path otherwise it finds other better path for reaching at the destination. When the forward ant successfully reaches the destination node, it inherits all the information and return from the destination node using the same path. Thus the pheromone value of that path will be increased.

Gianni Di Caro *et al.* [10] proposed that during the route discovery from source to destination except the destination node information all other field of the routing table will be updated. In the ant net algorithm, new ant net is created from source to destination. During finding the food the ant determine the next node has been visited before it shows that the ant is entering in the cycle and must die.

Gianni Di Caro *et al*. [11] proposed Anthocnet which is a hybrid multipath algorithm. In this algorithm, when a packet is received by the node, it check first of all in the routing table that the information of the destination node whether available or not. If the information of the destination node is available the node forwards the packet otherwise that node broadcast all the forward ants to finds the possible paths from source to destination.

Ayman M [12] proposed TARA protocol uses an objective trust model. In the TARA algorithm the trust value is consider for relative behavior of nodes which forward only small packets. In route discovery, the node find the destination node information is available or not in routing table. If the information is not found in the routing table the route ant agent is broadcasted.

Gurpreet Singh *et al*. [13] proposed ANTALG has considered the random selection of source and destination nodes and exchanges the agents between them. In the ANTALG algorithm source node randomly selects a node from the set of nodes and sends towards the randomly chosen destination.

Parimal Thulusiraman *et al*. [14] proposed ant colony optimization algorithm to determine the best path from any node to any other node in the network. When the packet is received, the ant chooses the node which creates the best path from source to destination. The ant selects the node which was not visited yet by any other ants. If no node exists then ant chooses the neighbor node by pheromone value.

## IV. CONCLUSION

Security in Mobile Ad-Hoc Network (MANET) is the main concern. So, in this paper we considered three different security approaches of MANET which are Bayesian Watchdog, Trust based system and Ant colony optimization. At the time in routing, when we need to establish secure routing from malicious node, the trust based system and Bayesian watchdog is used. While ant colony is used for discovering trusted shortest path and for optimality, as it is more prone to intruder attack, an alternate path which is secure from intruder attack would be developed because intruder would be interested in shortest path. A detailed analysis of various security approaches has also been provided. In future work we are going to implement these approaches and going to find complexity of each approach.

APPENDIX

Table.1: Various Security Approaches

| No. | Paper title | Year | Author | Parameters | Limitation |
|---|---|---|---|---|---|
| | **Bayesian Watchdog** | | | | |
| 1 | A Novel Approach for the Fast Detection of Black Holes in MANETs[6] | 2013 | Manuel D. Serrat-Olmos, Enrique Hernández-Orallo, Juan-Carlos Cano, Carlos T. Calafate, Pietro Manzoni | Global detection time, threshold, observation time, transmission, mobility pattern of the nodes | Slower detecting speed than collaborative watchdog approach |
| 2 | Collaborative Watchdog to Improve the Detection Speed of Black Holes in MANETs[22] | 2011 | Manuel D. Serrat-Olmos, Enrique Hernández-Orallo, Juan-Carlos Cano, Carlos T. Calafate, Pietro Manzoni* | Observation, reputation information, threshold, level of trust | The small amount of black holes, that are not detected with the bayesian atchdog, are now detected by the collaborative bayesian watchdog. |
| 3 | Mitigating Routing Misbehavior in Mobile Ad Hoc Networks[23] | 2013 | Sergio Marti, T.J. Giuli, Kevin Lai, and Mary Baker | Threshold bandwidth, time | It might not detect a misbehaving node in the presence of 1) ambiguous collisions, 2) receiver collisions, 3) limited transmission power, 4) false misbehavior, 5) collusion, and 6) partial dropping. |
| 4 | Black-Hole Attacks in P2P Mobile Networks Discovered through Bayesian Filters[24] | 2010 | Jorge Hortelano, Carlos T. Calafate, Juan Carlos Cano, Massimiliano de Leoni, Pietro Manzoni, and Massimo Mecella | Tolerance threshold, Fading value, Updating time. | There is influence of the noisy observation upon the accuracy. |
| 5 | Optimizing the efficiency of watchdog IDS in MANETs using selfishness information and bayesian filtering[25] | 2014 | Varsha Himthani, Prashant Hemrajani & Sachin Sharma | Lower Bound Value, Upper Bound Value, estimate of maliciousness, estimate of cooperativeness | There is influence of the noisy observation upon the accuracy. |
| 6 | Enhanced Intrusion Detection System for Discovering Malicious Nodes in Mobile Ad hoc Networks[26] | 2007 | Nidal Nasser and Yunfeng Chen | Threshold, time | If the real malicious node is on all paths from specific source and destination, then it is impossible for the source node to confirm with the destination of the correctness of the report. |
| | **Trust based system** | | | | |
| 7 | Multi-Path Trust-Based Secure AOMDV Routing in Ad Hoc Networks[15] | 2011 | Jing-Wei Huang, Isaac Woungang, Han-Chieh Chao, Mohammad S. Obaidat | Route selection time, trust compromise, simulaion time, data rate | During the XOR process if the pathe selected is not appropriate for transferring whole or a part of data than the process is restarted again |
| 8 | Trust-based on-demand multipath routing in mobile ad hoc networks[16] | 2010 | X. Li Z., Jia P., Zhang R. Zhang, H. Wang | Control packet forwarding ratio, Data packet forwarding ratio, Packet forwarding ratio | More routing packet overhead |
| 9 | Fuzzy Membership Function in a Trust Based AODV for MANET[17] | 2013 | Partha Sarathi Banerjee, J. Paulchoudhury, S. R. Bhadra Chaudhuri | Realiability, residual energy, buffer occupancy, packet generation rate, speed, packet delivery ratio, average end to end delay | More overhead |
| 10 | Light-weight trust-based routing protocol for mobile ad hoc networks[18] | 2010 | N. Marchang, R. Datta | Packet delivery ratio, Malicious packet drop ratio, End-to-end delay, Route frequency, Routing load, Average throughput | Some packets are dropped by malicious nodes initially to lean the behavious of malicious nodes |

| 11 | Trust Based QOS Protocol(TBQP) using Metaheuristic Genetic Algorithm for Optimizing and Securing MANET[19] | 2014 | Sherin Zafar, Prof. (Dr) M K Soni | End to end delay, throughput | Less efficiency |
|---|---|---|---|---|---|
| 12 | A trust-based framework for data forwarding in opportunistic networks[20] | 2011 | Na Li, Sajal K. Das | Data generation rate, message lifetime, simulation time | Not highly secured as the key to defend arbaritarily forwarding data in DTNs. |
| **Ant Colony Optimization(Bio-inspired)** | | | | | |
| 13 | HOPNET : A Hybrid ant colony optimization routing algorithm for Mobile Ad hoc Network[7] | 2008 | jianping Wang,Eseosa Osagie,Parimal Thulasiraman,Ruppa K Thulasiram | Maximum speed, simulation time, propagation time, data rate | Does not scale well for large networks |
| 14 | Implementation and evaluation of antnet a distributed shortest path algorithm[8] | 2005 | Mitthias Strobbe,Vincont Verstrate,Erik van Brensegem,Jan Coppons,Mario Pickavet,Piol Demeester | Packet size, elaboration time, pause time | long delays in propagating routing information |
| 15 | ANTNET: Distributed Stigmergetic Control for Communication Network[10] | 1998 | Gianni Di Caro,Macro Dorig | Packet size, simulation time,paus time, elaboration time, data rate | More overhead |
| 16 | AntHocNet:an Ant-based Hybrid Routing Algorithm for Mobile Ad-Hoc Network[11] | 2008 | Gianni Di Caro, Frederick Ducatelle and Luca Maria Gambardella | Maximum speed, simulation time, pause time, transmission range, data rate | More Overhead |
| 17 | TARA : Trusted Ant Colony Multi Agent Based Routing Algorithm for Mobile Ad-hoc Network[12] | 2008 | Ayman M Bahaa | Maximum speed, simulation time, pause time, transmission range, data rate | On increasing pause time the packet delivery ratio start increasing |
| 18 | ANTALG:An innovative ACO based Routing Algorithm for MANETs[13] | 2014 | Gurpreet Singh,Neeraj Kumar,Anil Kumar Verma | simulation time, pause time, propagation delay | Less packet dropped compare to HOPNET |
| 19 | A parallel ant colony optimization algorithm for all pair routing in MANET[14] | 2003 | Mohammad Towhidul Islam, Parimala Thulusiraman, Ruppa K Thulasiran | Packet size, simulation time, maximum speed | Cache pollution |